\documentclass[runningheads,a4paper]{llncs}
\usepackage{etex}
\reserveinserts{28}
\pdfoutput=1

\usepackage[linesnumbered,lined,boxed,commentsnumbered,ruled]{algorithm2e}

\SetAlFnt{\small}
\SetAlCapFnt{\small}
\SetAlCapNameFnt{\small}
\SetAlCapHSkip{0pt}
\IncMargin{-\parindent}
\usepackage{ textcomp }
\usepackage{ comment }
\usepackage[color=yellow!60,textsize=footnotesize,obeyDraft,draft]{todonotes}
\usepackage{colortbl}
\usepackage{booktabs}
\usepackage{capt-of}
\usepackage{makecell}
\usepackage{textcomp }
\usepackage{multirow}
\usepackage{listings}
\usepackage{graphicx}
\usepackage{colortbl}
\usepackage{booktabs}
\usepackage{amsmath}
\usepackage[font=footnotesize,labelfont=bf]{subcaption}
\usepackage[hidelinks]{hyperref}
\usepackage[capitalize]{cleveref}
\usepackage{tikz}
\usepackage{pgfplots}
\usepackage[para,online,flushleft]{threeparttable}
\usepackage{multirow}
\usepackage{wasysym}
\usepackage{amssymb}
\usepackage{enumitem}
\usetikzlibrary{arrows}
\usepackage{subcaption}
\usepackage{breakcites}
\newcounter{mnotei}
\newcolumntype{L}[1]{>{\raggedright\let\newline\\\arraybackslash\hspace{0pt}}m{#1}}
\newcolumntype{C}[1]{>{\centering\let\newline\\\arraybackslash\hspace{0pt}}m{#1}}
\newcolumntype{R}[1]{>{\raggedleft\let\newline\\\arraybackslash\hspace{0pt}}m{#1}}

\newcommand{\includegraphicsmaybe}[2]{
    \IfFileExists{#2}{\includegraphics[#1]{#2}}{
    \detokenize{File #2 is missing, maybe you need to run plots.py?}
}}

\begin{document}
\bibliographystyle{ieeetr}
\mainmatter

\title{The IoT energy challenge: A software perspective}

\titlerunning{The IoT energy challenge: A software perspective}

\author{Kyriakos Georgiou\inst{1}, Samuel Xavier-de-Souza\inst{2}, Kerstin Eder\inst{1}}

\authorrunning{K. Georgiou et al.}
\institute{University of Bristol, UK \and Universidade Federal do Rio Grande do Norte, Brazil}
\tocauthor{Authors' Instructions}
\maketitle

\makeatletter
\renewcommand\subsubsection{\@startsection{subsubsection}{3}{\z@}%
                       {-18\p@ \@plus -4\p@ \@minus -4\p@}%
                       {4\p@ \@plus 2\p@ \@minus 2\p@}%
                       {\normalfont\normalsize\bfseries\boldmath
                        \rightskip=\z@ \@plus 8em\pretolerance=10000 }}
\makeatother

\begin{abstract}
The Internet of Things (IoT) sparks a whole new world of embedded applications. Most of these applications are based on deeply embedded systems that have to operate on limited or unreliable sources of energy, such as batteries or energy harvesters. Meeting the energy requirements for such
applications is a hard challenge, which threatens the future growth of the IoT. Software has the ultimate control over hardware. Therefore, its role is significant in optimizing the energy consumption of a system. Currently, programmers have no feedback on how their software affects the energy consumption of a system. Such feedback can be enabled by energy transparency, a concept that makes a program's energy consumption visible, from hardware to software. This paper discusses the need for energy transparency in software development and emphasizes on how such transparency can be realized to help tackling the IoT energy challenge.
\end{abstract}

\section{Introduction}

The IoT is no longer just a buzzword in the media; it is becoming a reality. The emergence of IoT led us into a new era of innovation and creativity. This sets high expectations on both the research community and industry for delivering the necessary technological advancements, that will allow for the materialization of new IoT applications.

Powering billions of embedded devices deployed into the environment is one of the biggest challenges that IoT faces. Battery-based solutions tend to be impractical and costly due to the need of recharging or replacement. Energy harvesting appears as a viable option for many IoT applications, but it comes with two caveats. Firstly, it is often an unreliable source of energy. Secondly, there is still a large gap between the energy it can deliver and the required energy budget for many IoT applications. For mission critical IoT applications, such as health-care, completing a task before running out of energy budget is vital.

Traditionally, hardware innovation has been the safe heaven to achieve sufficiently large savings of energy in Information and Communication Technology (ICT). Similarly, hardware innovation is currently the prominent response to tackle the energy challenge IoT faces. New ultra-low-energy embedded devices were introduced, and existing technologies were customized to create new, more energy efficient versions, such as the Bluetooth Low Energy (BLE). These are well suited for energy-critical applications. But, is that all we can do to tackle this challenge? 

It is estimated that up to 80\% of the total energy consumption of an embedded system is due to software-related activities~\cite{Luo:2009}. Inefficient software can drive energy-efficient hardware to waste the system's energy budget. Steve Furber, principal designer of the ARM microprocessor, gave an interview in 2010~\cite{2010:CSF:1716383.1716385} and stated:

\begin{quotation}
\small
\emph{``Programmers will not be able to afford to be ignorant about the energy cost of the programs they write ... You need tools that give you feedback and tell you how good your decisions are. Currently the tools don't give you that kind of feedback.''}
\end{quotation}

These tools are now needed more than ever to overcome the IoT energy challenge. Programmers have very limited information on how much energy their programs consume, and which parts use the most energy. This has two important implications for the development of IoT applications:

\begin{enumerate} 
\setlength\itemsep{0.2em}
 \item Much guesswork is needed, and thus bad energy-related choices are only identified at a late stage when the system malfunctions due to a failure to meet the system's energy requirements. 
 \item The high level of expertise needed and the lack of energy-aware development tools  significantly reduce the number of embedded developers who are able to deliver energy-critical systems.
\end{enumerate}

Because of this, the whole process of developing energy-constrained applications becomes difficult and costly. There is a need for tools that expose the software's effect on the energy consumption of a system. Such energy transparency will allow programmers, toolchains and runtime systems to make energy-aware decisions in order to meet the strict energy constraints of the IoT. 

\section{Enabling energy transparency is difficult}
\label{sec:transparencyIssues}

Various layers have been introduced in the system stack, abstracting away complex details to make programming easier. This prevents software developers from understanding the impact of their coding choices on the way hardware is utilized at runtime. The task of enabling the required energy transparency can be seen as equivalent to the task of reverse engineering the code transformations that take place between and at each software abstraction layer. This will allow the propagation of resource usage information from hardware to software. Such a reverse engineering process is a hard challenge and highly dependent on the architecture and compiler choice. Instead, novel energy transparency techniques are needed which can approximate the energy consumption of a program at different software abstraction levels, without the need for reverse engineering. For these techniques to enable energy-aware software development, the following requirements must be met:

\begin{enumerate}[leftmargin=*]
\item{\textbf{Both the actual energy consumption and bounds must be provided:}} Energy consumption bounds will guide developers in meeting strict energy budget requirements. Actual energy consumption estimates are necessary to serve as a benchmark for potential energy optimizations.
\item{\textbf{Energy transparency at multiple levels of software abstraction:}} Software developers and toolchains can significantly influence the energy consumption of a program mainly at three software abstraction levels; the source code, the compiler's Intermediate Representation (IR), and the Instruction Set Architecture (ISA).
\item{\textbf{Fine-grained energy characterization of software:}} Identifying energy hot-spots 
requires the ability to attribute the energy consumption estimates to the various basic software components, such as Control Flow Graph (CFG) basic blocks. 
\item{\textbf{Target and programming language independence:}} Target and programming language agnostic techniques must be provided in a common framework to enable practical and cost effective energy-aware development for a large number of embedded architectures. 

\item{\textbf{The multi-threaded and multi-core case must be considered and explored:}} Novel multi-core, multi-threaded embedded architectures emerged over the last decade, driven by the increasing demand for more computing power. As this trend is expected to grow in the future, it is important to consider such architectures. Also, they must be explored for parallel codes because they offer potentially large energy savings when the number of cores increases and the cores' voltage and frequency decrease.

\item{\textbf{Enable design space exploration:}} Developers and toolchains need to apply multi-objective optimizations to find the optimum balance between the available resources, such as execution time, energy, code size, number of cores and threads used. To enable this, energy transparency techniques have to provide sufficiently accurate feedback on the effect that each different configuration has on the resources of interest.
\item{\textbf{Fast and easy to deploy:}} Energy consumption estimation speed is critical to the iterative software optimization process. Furthermore, energy transparency techniques should be easy to deploy and use. 
\end{enumerate}

\section{Existing approaches and limitations}

This section examines the state of the art of energy transparency techniques and their limitations in regards of the above requirements.

\subsection{Measuring energy consumption}
While physical measurements are potentially most accurate to determine the energy consumption of a program, they fail to meet many of the requirements set in~\Cref{sec:transparencyIssues}. Firstly, measurements typically require sophisticated equipment and hardware knowledge that most software developers lack. Secondly, most hardware components have no provisions for energy measurements; thus custom modifications are needed to probe their power supply. This makes physical measurements difficult to deploy and use. Furthermore, extracting energy consumption bounds with end-to-end measurements is inappropriate in most cases, as the whole input space would need to be exhaustively searched. Finally, fine-grained energy characterization of software can be challenging. Usually, this requires expensive measuring equipment such as oscilloscopes, that can support a high sampling frequency.

\subsection{Estimating energy consumption}
\label{energyEstimation}

Estimating the energy consumption of a program for a particular hardware platform requires two main elements: an analysis technique and a way to convey energy information to the analysis. The latter is typically done via energy modeling. An energy model statically captures the dynamic behavior of a processor in regards to its energy consumption characteristics; for example, it can associate energy costs with atomic units in a program, such as ISA instructions, or to various events, such as a cache miss.

\begin{figure}[ht]
\centering
  \includegraphics[width=1\textwidth,trim=0.1cm 9.6cm 4.4cm 0.3cm,clip]{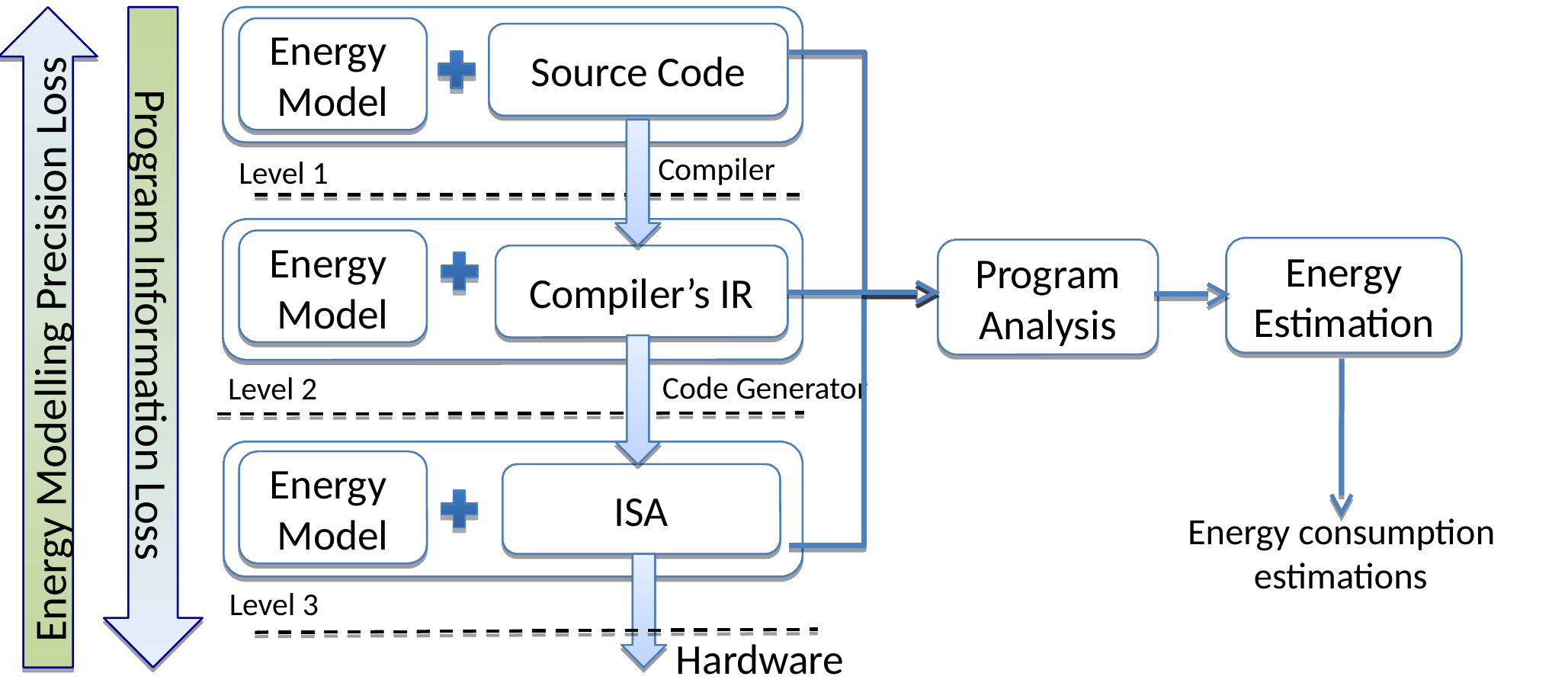}
\caption{Software abstraction level and energy estimation accuracy trade-off.}
\label{fig:analysisLevels}
\end{figure}

To perform energy estimation at a software level of abstraction, an energy model is required at the same level. An energy analysis technique automatically inherits any precision loss existing in the energy model that the technique utilizes. \Cref{fig:analysisLevels} demonstrates the trade-off between the energy consumption estimation accuracy and the level of software abstraction. 
Modeling at a lower level is always more accurate as it is closer to the hardware, where the actual power dissipation occurs. On the other hand, when moving to higher levels of abstraction, the amount of program information, such as types and loop structures, increases. Such information can be crucial for static code analysis and optimization. However, when moving from source code to the ISA level, much of this information is lost due to the various transformation and optimization passes that occur at the respective levels of the software stack.

In the next sections, the state of the art of energy consumption estimation techniques is examined.

\subsubsection{Profiling-based energy consumption estimation}

In this case, estimation is performed by collecting execution statistics and utilizing them with an appropriate energy model. Such a model needs to provide energy information for the various entities that occur in the execution statistics. Three main techniques are used to collect execution statistics for energy consumption estimation:

\emph{\textbf{Simulation:}} Typically performed at low hardware design levels, such as the 
Register-Transfer Level (RTL)~\cite{Ye:2000}, thus it is difficult to achieve fine-grained energy consumption attribution to the various software components. Moreover, modeling and profiling at such low levels is impractical for most commercial embedded processors since essential circuit information, such as the effective capacitance of major architectural blocks, is not available. 
To account for these issues, simulation-based energy estimation has been performed at the ISA level~\cite{Tiwari1996}. Energy models at this level can be constructed for deeply embedded commercial processors by treating the hardware as a black box~\cite{Tiwari1996}. ISA energy models are less accurate than lower-level models, but ISA simulation is considerably faster than hardware simulation and allows for fine-grained energy characterization of software. 

Energy modeling and profiling at the ISA level are insufficient for more complex architectures or system-level energy consumption estimation. This is because it is difficult to statically capture the behavior of performance-enhancing hardware components, such as caches.

Generally, simulation-based energy consumption estimation tends to be slow. This makes it difficult for the technique to be incorporated in software development tools, were instant feedback is required for an iterative process of optimizing energy consumption. 

\emph{\textbf{Code Instrumentation:}} This is performed by instrumenting the code with instructions that extract execution statistics at runtime. The main challenge is to extract the statistics out of the hardware and to minimize the overhead of the instrumentation that can significantly impact the estimation's accuracy. Therefore, the amount of execution statistics collected are typically less compared to simulation. This makes the retrieved energy consumption estimations less accurate than simulation-based estimations. A major advantage of this method is that it is significantly faster than simulation-based energy estimation. 

Recent work~\cite{Georgiou:2017} demonstrated a new profiling technique that collects execution statistics at the compiler's IR level. This was combined with a dynamic mapping technique that  lifts an ISA energy model to the compiler's IR, to retrieve energy estimations at the IR level. The technique guarantees no energy overhead in the estimations due to instrumentation code, and achieved an average accuracy of 2.5\%.

\emph{\textbf{Performance Monitoring Counters (PMC):}} Statistical PMC-based estimation is preferable for more complex architectures were ISA-level modeling and analysis is insufficient to capture their complexity. PMC execution statistics can be used to construct energy models and estimate the energy consumption of multi-threaded/core architectures~\cite{Schubert:2012}. Run-time power estimation can be enabled using PMC~\cite{xscaleunitevents}. This allows for energy-aware decisions to be made at runtime.

For a given processor, energy modeling and profiling using PMCs can be challenging due to
the limited types of the events that can be monitored and the restricted number of counters that
can be sampled simultaneously. The same constraints apply when trying to port the PMC-based
modeling and estimation techniques to a new target. Moreover, runtime use of such PMC-based
energy estimation methods could cause a significant overhead on the system's performance~\cite{Schubert:2012}.

\subsubsection{SRA-based energy consumption estimation}

Similar to physical measurements, it is impractical to capture energy consumption bounds using profile-based estimation. Static Resource Analysis (SRA) offers a better alternative. The following two techniques have been used for statically estimating energy consumption bounds:

\emph{\textbf{Automatic Complexity Analysis:}} This has been used to create cost relations that capture energy in terms of program-input size at both the ISA and the compiler's IR levels~\cite{isa-vs-llvm-fopara}. The technique can be fully automatic and programming-language independent. The main drawback is the difficulty to extract a closed-form solution for the cost relations of a program~\cite{Georgiou:2017}. Therefore, the approach does not scale well to large programs with complex structure.

\emph{\textbf{Implicit Path Enumeration Technique (IPET):}} This is the most popular method for Worst Case Execution Time (WCET) analysis~\cite{Wilhelm:2008}. In~\cite{Jayaseelan2006}, the technique was used to extract energy consumption bounds on a simulated processor. More recently~\cite{Georgiou:2017}, IPET was applied at the ISA level of a multi-threaded/core embedded architecture, using an ISA energy model, and at the compiler's IR level, using a mapping technique that lifts ISA energy models to the compiler's IR level. The authors also demonstrated how the technique can be used for design space exploration for two concurrency patterns: task-farms and pipelined programs. 

Like WCET estimation, SRA-based energy estimation works best with predictable architectures and software. Using, SRA to analyze multi-threaded programs with complex communication patterns is a hard challenge. Furthermore, currently, there is no practical method to perform average-case static analysis~\cite{townley2013practical}. 

\section{Outstanding challenges}

Techniques that are based on energy modeling, code-instrumentation and SRA avoid the need for difficult to deploy physical energy measurements and expensive simulations. Moreover, they can provide fine-grained energy characterization of software at multiple levels of abstraction with good precision. A combination of SRA and code-instrumentation techniques provides both the bounds and the actual energy consumption. Furthermore, new energy estimation techniques, such as those presented in~\cite{Georgiou:2017}, are compiler and architecture agnostic, provided an ISA energy model exists. Therefore, such techniques can be relatively easily integrated into development toolchains to provide feedback-directed energy optimization. However, there is still a number of outstanding challenges that need to be addressed. 

Currently there is no practical solution to provide tight upper energy consumption bounds. SRA approaches combined with worst case energy models can lead to significant overestimation~\cite{Jayaseelan2006}. Symbolic simulation at the RTL could retrieve tighter bounds,~\cite{Cherupalli:2017}, but such approaches require sensitive architectural information, typically not available for commercial processors. To retrieve tight energy bounds at the ISA level, data-sensitive energy models and static analysis would be required. These have to take into account the inter-instruction effects, caused by the operand values used for each instruction, and identify the worst case data input. Recent work demonstrated that finding the data that will trigger the worst case energy consumption is an NP-hard problem and that no practical method can approximate (within reasonable time in general) tight energy consumption upper bounds within any level of confidence~\cite{DBLP:journals/corr/MorseKE16}.

Energy models are typically characterized while using a constant power source. This is an ideal condition, as there are no significant variations in the power supplied to the processor. This is not an issue when the energy estimations retrieved are used to optimize the energy consumption at development time. However, IoT applications typically run on a battery or on an energy harvester. To be able to use the energy estimations for making real-time decisions, the power profile of the power source for a specific application has to be taken into consideration by both the energy modeling and the energy consumption estimation techniques.

Activities external to the processor, such as sensing and communicating, can consume significantly more energy than the computation part of an embedded system. Computation usually controls such activity. Therefore, the energy usage of peripherals and I/O operations can be profiled and included as part of the energy cost of the computation that triggers them. This will allow existing energy estimation techniques to provide system-wide energy estimates. Such an approach is more feasible for systems with predictable behavior.

While existing SRA-based estimation techniques can handle the complexity of deeply embedded architectures, they generally do not scale well to new multi-core/threaded architectures. Concurrent software introduces complexities over traditional sequential variants, which SRA is inherently limited to cope with. The task is even harder than estimating WCET since any computation contributes to the worst-case energy scenario, while, for WCET analysis, only the computation that causes the worst case needs to be considered. Therefore, SRA support for multi-threaded and multi-core software is currently limited to a range of simpler concurrent software patterns. Both hardware and toolchain vendors need to work closely together to provide novel architectures and methods that will allow energy transparency by design.

The inter-thread core activity must be considered, to enable the interplay between performance and energy on many-, multi-core and multi-threaded architectures that support voltage and frequency scaling. Finding optimal configurations based on runtime execution statistics and a power model for the architecture is more practical, rather than using simulation. This is because simulation is typically several orders of magnitude slower than hardware execution and thus difficult to support an interactive optimization software development process. Furthermore, an energy modeling that accounts for voltage and frequency scaling and parallelism can answer optimization questions such as how many cores and at what frequency they should run for a given performance target\cite{barros2015optimal,hager2016exploring}. Hardware vendors should provide more support to enable such techniques; for example more PMCs. Finally, the combination of both static and runtime analysis must be explored to realize energy savings beyond those  than  can be achieved when these approaches are used independently.

Perhaps, the biggest challenge of enabling energy-aware software development lies in the tool-vendor's and programmer's perception. Traditionally, optimizing energy consumption has been treated as a side effect of improving execution time. Toolchains have been long focused on improving execution time, and developers are left assured that a compiler will do the best possible job on optimizing their code for both time and energy. The IoT energy challenge is an opportunity to start treating energy consumption as a first class citizen while developing software.
\section{Conclusion}

Hardware offers energy saving capabilities for the software to exploit. The responsibility lies on the system engineer to program and configure the selected device in the most energy efficient way for the task at hand. Development toolchains need to be enhanced with energy transparency that provides the necessary feedback to enable energy-aware software development. Hardware vendors need to provide architectures and information, such as energy models, that will support energy transparency. Finally, tool vendors and programmers need to take a more active role in delivering energy efficient systems. We can't keep ignoring the software role in the energy consumption of a system.

\bibliography{typeinst}

\end{document}